\documentclass[useAMS,usenatbib]{mn2e}

\usepackage{graphicx}	
\usepackage[british,UKenglish,USenglish,english,american]{babel} 
\usepackage{epstopdf}

\title[A comparative analysis of white dwarf cooling sequences]{A comparative analysis of the observed white dwarf cooling sequence from globular clusters}
\author[F. Campos et al.]{Fab\'iola~Campos$^1$\thanks{fabiola.campos@ufrgs.br} P.~Bergeron$^2$, A.D.~Romero$^1$, S. O.~Kepler$^1$,  G. Ourique$^1$,
\newauthor J. E. S.~Costa$^1$, C. J.~Bonatto$^1$, D. E.~Winget$^3$, M. H.~Montgomery$^3$, T. A.~Pacheco$^1$,
 \newauthor  L. R.~Bedin$^4$\\
 $^1$Departamento de Astronomia, Universidade Federal do Rio Grande do Sul, 
Av. Bento Gon\c{c}alves 9500 Porto Alegre 91501-970, RS, Brazil\\
  $^2$D\'epartement de Physique, Universit\'e de Montr\'eal, C.P. 6128, 
Succursale Centre-Ville, Montr\'eal, Qu\'ebec H3C 3J7, Canada\\
  $^3$Department of Astronomy, University of Texas at Austin, Austin, TX, USA\\
  $^4$Istituto Nazionale di Astrofisica, Osservatorio Astronomico di Padova, 
Vicolo dell' Osservatorio 5, Padova, IT-35122, Italy\\}
\begin{document}

\date{Accepted 2015 December 8.  Received 2015 December 3; in original form 2015 August 11}

\pagerange{\pageref{firstpage}--\pageref{lastpage}} \pubyear{2015}

\maketitle

\label{firstpage}

\begin{abstract}
We report our study of features at the observed red end of the white dwarf cooling sequences for three Galactic globular clusters: NGC\,6397, 47\,Tucanae and M\,4.  We use deep colour-magnitude diagrams constructed from archival Hubble Space Telescope (ACS) to systematically investigate the blue turn at faint magnitudes and the age determinations for each cluster. We find that the age difference between NGC\,6397 and 47\,Tuc is 1.98$^{+0.44}_{-0.26}$\,Gyr, consistent with the picture that metal-rich halo clusters were formed later than metal-poor halo clusters. We self-consistently include the effect of metallicity on the progenitor age and the initial-to-final mass relation.  In contrast with previous investigations that invoked a single white dwarf mass for each cluster, the data shows a spread of white dwarf masses that better reproduce the shape and location of the blue turn. This effect alone, however, does not completely reproduce the observational data - the blue turn retains some mystery.  In this context, we discuss several other potential problems in the models.  These include possible partial mixing of H and He in the atmosphere of white dwarf stars, the lack of a good physical description of the collision-induced absorption process and uncertainties in the opacities at low temperatures. The latter are already known to be significant in the description of the cool main sequence. Additionally, we find that the present day local mass function of NGC\,6397 is consistent with a top-heavy type, while 47\,Tuc presents a bottom-heavy profile. 
\end{abstract}


\section{Introduction}
Galactic globular clusters are among the largest and oldest stellar systems in the Milky Way. They contain thousands to millions of stars at approximately the same distance from the Sun: their cluster half-light radii are $\rm\sim4\,pc$ \citep[e.g.][]{bonatto07} while their distances are larger than $\rm2\,kpc$ \citep{harris96}. The stars forming the globular clusters are considered excellent laboratories to study stellar evolution. They are assumed to be part of a simple stellar population, meaning that all stars have been formed at essentially the same time, from the same cloud and have the same initial chemical composition \citep{moehler08}. More precise investigations have shown that most of the globular clusters are not composed of a simple stellar population; instantaneous star formation and complete chemical homogeneity of the original cloud should not be expected \citep[e.g.][]{dantona05, piotto15}. An extreme case is that of $\omega$ Centauri, which clearly has multiple main sequences \citep{bedin04}
and a double white dwarf cooling sequence \citep{bellini13}. However, such
features are not so extreme. Nevertheless, globular clusters remain excellent laboratories, composed of stars from the central hydrogen-burning limit in the main sequence to cool white dwarf stars. 

The white dwarf phase is the evolutionary endpoint of stars with initial masses lower than 
roughly 10\,$M_{\odot}$ \citep[e.g.][]{ibeling13, doherty15}.  Thus they represent the collective fate of $\sim$97\% of the stars \citep[e.g.][]{fontaine01}. \cite{moehler08} argue that white dwarf stars in clusters offer an advantage over field white dwarf stars in that they provide the opportunity to constrain the initial-to-final mass relation. Besides, most of the globular clusters are mono-metallic systems with respect to iron, with a minimum spread in age. Moreover, the local densities of the halo, thick, and thin disk white dwarf populations are lower than their local density in a globular cluster, making it possible to observe large samples of white dwarfs without the need for wide-field surveys. There are several substantial obstacles in the study of white dwarf stars in globular clusters: they are more distant than the neighbouring stars, the effect of crowding is significant and white dwarf stars are very faint, requiring deep photometric observations with high quality. In that context, \cite{hansen98} pointed out that only when it were possible to examine the faint white dwarf stars in globular clusters it would be possible to empirically test the effects of advanced age and low-metallicity progenitors. The globular clusters NGC\,6397, 47\,Tuc and M\,4 are the first ones that have data reaching the red end of the white dwarf cooling sequence, thus providing the means to study stellar evolution and the coolest observable white dwarf stars in globular clusters in an unprecedented way.

The pioneering results achieved by \cite{richer06} with deep HST/ACS of NGC\,6397 showed, for the first time, the colour-magnitude diagram of a globular cluster down to the main sequence hydrogen-burning limit, and pointed the existence of a red cut-off in the white dwarf cooling sequence at $m_{F814W}$=27.8. Using artificial star tests and star-galaxy separation, \cite{hansen07} demonstrated that the cut-off represents a real truncation of the white dwarf stars luminosity function and estimated the age of NGC\,6397 to be $11.47\pm 0.47\,$Gyr. More recently, with the same method, \citet{hansen13} determined the age of the metal rich globular cluster 47\,Tuc as being $9.90\pm 0.70\,$Gyr, finding that this cluster is approximately 2\,Gyr younger than NGC\,6397. However, \citet{garcia-berro14} determined an age of $\sim12.00\,$Gyr for 47\,Tuc, also using the cooling sequence method but with different models.  This brought up questions about the reliability of ages determined through the white dwarf cooling sequence \citep{forbes15} and the importance of the theoretical evolutionary models used.

When \cite{hansen07} compared their best fit model to the data, they noticed that the only features of the observations that were not well reproduced by their best fit models were the blue colours at magnitudes around F814W=27.25, indicating a mismatch between theoretical and true colours at the faintest temperatures (T$_{\rm eff}\sim\,5\,000\,K$). They argued that the mismatch between models and data indicated either residual deficiencies in the models or that the atmospheres could be composed of a mixture of hydrogen and helium. They also proposed that the blue turn in the colours was driven by collision-induced absorption of molecular hydrogen,  a significant effect on environments composed of molecules or dense, neutral and non-polar atoms \citep{bergeron95,hansen98,saumon99,borysow00}. For non-polar molecules, collision-induced absorption may become the dominant source of opacity over a wide range of the infrared part of the spectrum. Hydrogen and helium are the most abundant atoms in stellar atmospheres but helium does not form molecules and $\rm H_2$ is non-polar; therefore, both collision-induced absorption involving H and He ($\rm H_2-H_2$ and $\rm H_2-He$), and molecules composed of less abundant elements (C, N, O, Ti, and others), if present, dominate the opacity in cool stars.

On the other hand, it is well known \citep[e.g.][]{hurley03, bedin05, bedin10, salaris10}, that the blue hook seen in white dwarf cooling sequences of clusters is mainly caused by a spread in mass. In a star cluster, the massive white dwarf stars form first from the most massive progenitors and, as time goes by, the less massive white dwarf stars only get to the top of the cooling sequence. Also, the more massive the white dwarf star is, the smaller it is and the slower it cools - until crystallisation. So, in a star cluster, as the star evolves, the white dwarf cooling sequence becomes redder in the colour-magnitude diagram, and the less massive white dwarf stars reach the luminosity of the more massive ones, causing the blue turn at the bottom of the cooling sequence. 

With HST photometry, \cite{bedin05,bedin08a,bedin08b,bedin10,bedin15} reached the end of the white dwarf cooling sequence of the open clusters NGC\,6791, NGC\,2158 and NGC\,6819, that also presented the blue turn feature. For NGC\,6791, they determined a discrepancy between the age from the white dwarf stars (6\,Gyr) and that from the main sequence turnoff (8\,Gyr). \cite{garcia-berro10} argued that, if only the carbon-oxygen phase separation is taken into account, the age of NGC\,6791, determined by white dwarf cooling sequence would increase to only $6.4\pm0.2$\,Gyr, in disagreement with the age determinations by the main sequence turn-off ($8-9$\,Gyr, \citealt{bedin05}). But, if they considered a combination of $^{22}$Ne sedimentation and carbon-oxygen phase separation \citep{deloye02,bedin08a,bedin08b}, the white dwarf age corresponded to $8.0\pm0.2$\,Gyr \citep{garcia-berro10}. They also argued that the blue turn in the white dwarf cooling sequence is caused by the most massive white dwarf stars of the cluster, in agreement with \cite{hurley03}.  

The blue turn feature observed in NGC\,6397, and also detected in the globular clusters M\,4 and 47\,Tuc \citep{bedin09,kalirai12}, has been discussed by \cite{hansen07, richer13}, in terms of collision induced absorption, instead of a mass effect. 

The truncation of the white dwarf cooling sequence is a very sensitive age indicator as it is the limit at which most white dwarfs can have cooled in the lifetime of a cluster \citep{bedin09}, demanding an accurate description of this feature in order to obtain precise determinations of the ages of the clusters through the white dwarf cooling sequence. 

Our main goal is to investigate the white dwarf cooling sequences of globular clusters. We use our isochrone models of the white dwarf cooling sequences calculated with metallicities consistent with the ones from each cluster and fix the initial-to-final mass function to the metallicity dependent ones from \citet{romero15}. Thus, we determine the distance modulus, reddening, age and the parameter $\alpha$ from the present day local mass function for NGC\,6397 and 47\,Tuc. 

This paper is organised as follows: in Sect. \ref{data} we give a brief description of the photometric data used in our paper. In Sect. \ref{models} we describe the evolutionary models and the white dwarf cooling sequences we used to build our white dwarf isochrones. In Sect. \ref{analysis} we show our analysis and in Sect. \ref{discussion} we discuss the results obtained. Concluding remarks are given in Sect. \ref{conclusions}.

\section{Data}
\label{data}

We perform our analysis based on the photometric data of the white dwarf stars belonging to the globular clusters NGC\,6397, 47\,Tucanae and M\,4 (Fig. \ref{3clus}), obtained with the {\it Advanced Camera for Surveys} (ACS) of the {\it Hubble Space Telescope} ({\it HST}). These clusters are the only globulars to date with available data reaching the red end of the white dwarf cooling sequence, showing the turn towards bluer colours for the coolest white dwarf stars.

\subsection{NGC\,6397}
The deep photometry of NGC\,6397 was obtained  by \cite{richer06} as part of a large observation program in {\it HST} Cycle 13 (GO-10424, PI: H. Richer) totalling 126 orbits. The data consists of 252 exposures (179.7\,ks) in the F814W filter and 126 exposures (93.4\,ks) in F606W. After they obtained the photometry, \cite{richer06} applied image spread tests to determine which sources were stellar (star-galaxy separation). To obtain the proper motion cleaned colour-magnitude diagram, \cite{richer06} determined the displacements between the 2005 ACS data and the WFPC2 images taken in 1994 and 1997, centred at the same coordinates, with respect to member stars, and with this, the zero point of motion of the cluster. All stars lying below the 2$\sigma$ error box in the proper motion distribution in each magnitude were considered cluster members \citep{anderson08}. 

\subsection{47\,Tucanae}
For 47\,Tuc our analysis was based on the photometric data obtained with 121 orbits in {\it HST} Cycle 17 (GO-11677, PI: H. Richer). \cite{kalirai12} describe the observations as 117 long exposures in F606W (163.7\,ks) and 125 in F814W (172.8\,ks). The observations were centred at about 6.7\,arcmin west of the cluster centre which had previous data observed with ACS, making it possible to obtain the proper motion separation, in a similar way to the study made by \cite{anderson08} for NGC\,6397.

\begin{figure}
\resizebox{\hsize}{!}{\includegraphics[clip=true]{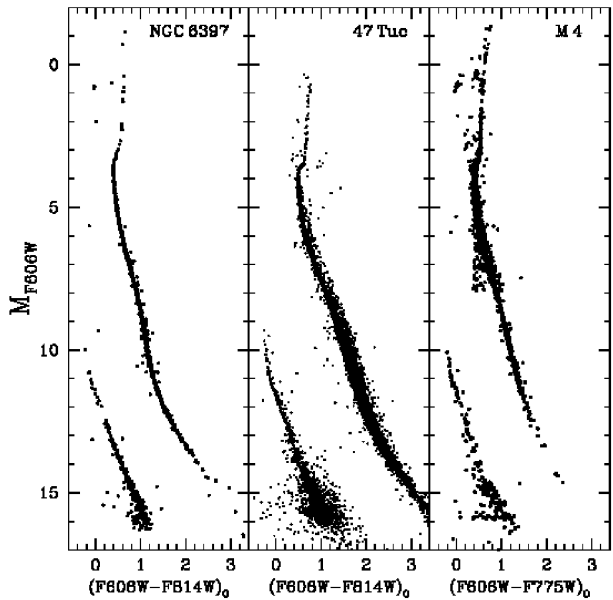}}
\caption{\protect
Proper motion cleaned colour-magnitude diagrams NGC\,6397, 47\,Tuc and M\,4. The axes in this plot are absolute magnitude and intrinsic colour in F606W. The data in this plot has proper motion correction for all clusters. Apparent distance modulus and reddening applied for the clusters are those from \citet{richer13} and references therein. 
}
\label{3clus}
\end{figure}

\subsection{M\,4}
\cite{bedin09} were able to reach the end of the white dwarf cooling sequence, at m$_{F606W}=28.5\pm$\,0.1, with 14 {\it HST} orbits. The M\,4 data was obtained in different programs, consisting of 20 exposures in the F606W filter (10 orbits) and 4 exposures in F775W (2 {\it HST} orbits) as part of the program GO-10146 (PI: L. Bedin) in {\it HST} Cycle 13, and F775W filter data from {\it HST} Cycle 11 (GO-9578, PI: J. Rhodes), consisting of 10 exposures of 360 s (2 {\it HST} orbits). The program GO-10146 also obtained short exposures in both F606W and F814W filters, and F606W archival material from program GO-10775 (PI: Sarajedini). Proper motions were measured with a technique similar to that described in \cite{bedin03,bedin06}, also considering the zero point of the motion of the cluster.

\section{Evolutionary Models}
\label{models}

White dwarf evolutionary models used in our analysis are those computed by \cite{romero15} for different metallicities. These models are the result of computations employing the {\small LPCODE} evolutionary code, the same code used by \citet{althaus05, althaus15} and \citet{renedo10}. They computed the evolution from the Zero Age Main Sequence, through the hydrogen and helium central burning stages, the full thermally pulsing and mass loss stages on the Asymptotic Giant Branch, and finally through the white dwarf cooling curve. 

The models also incorporate the effects of possible residual hydrogen burning present in 
low-metallicity progenitors, for which the mass loss is less efficient and as a 
result, the thickness of the hydrogen layer left on the white dwarf model increases 
with decreasing metallicity. Then, the larger the hydrogen mass on the envelope of 
the white dwarf the more efficient the residual burning at high effective temperatures 
on the cooling sequence \citep{renedo10, bertolami13}.

Here, we briefly mention some of the main input physics relevant for this work. A 
detailed description can be found in \cite{althaus05} and  \cite{romero15}. During the white dwarf cooling, the code  considers element diffusion due to gravitational settling, chemical and thermal diffusion \citep{althaus03}. At effective temperatures below $10\,000$K on the cooling curve, the outer boundary conditions are derived from non-grey atmosphere models from \cite{rohrmann12} which include the tail of Ly$\alpha$ in the optical  \citep{kowalski07}. 
The release of latent heat and gravitational energy due to carbon-oxygen phase separation is included following the \cite{horowitz10} phase diagram, consistent with the observations by \cite{winget09,winget10} and the massive pulsating white dwarf stars studied in \cite{romero13}. 

\begin{figure}
\resizebox{\hsize}{!}{\includegraphics[clip=true]{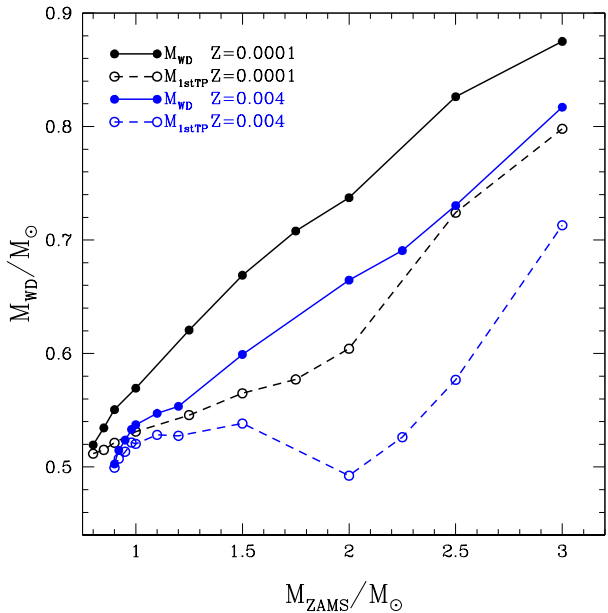}}
\caption{\protect
Initial - Final mass relation from \citet{romero15} considering the  final mass as the  mass of the hydrogen  free core at the first thermal  pulse (hollow circles) and as the  mass of the white dwarf at  the cooling curve (full circles). Black and blue curves correspond to sequences with initial metallicity Z=0.0001 and Z=0.004, respectively.}
\label{masstwd}
\end{figure}

We consider main sequence masses from $\sim$0.8 to $\sim$6.0 $M_{\odot}$, where the low mass limit depends on metallicity. The resulting white dwarf models show stellar mass values from $\sim$0.50 to $\sim$1.00$\,M_{\odot}$. Note that this mass interval corresponds to carbon-oxygen core stars. We consider the mass of the white dwarf as the actual final mass in the cooling curve, not the approximation of the mass of the hydrogen free core at the first thermal pulse. At the end of the AGB stage, when the star becomes a thermally pulsing TP-AGB star, the mass of the hydrogen free core will increase with each thermal pulse. \cite{kalirai14} estimated a growth of the core mass between 10\% to 30\% for a progenitor of 1.6-2.0$M_{\odot}$. Therefore the mass at the cooling curve is larger than the core mass at the first thermal pulse, leading to a different initial-to-final mass prescription. Figure \ref{masstwd} shows the initial-to-final mass relation resulting from \cite{romero15} computations for  Z=0.0001 and Z=0.004. With filled circles we depict the initial-to-final mass with a white dwarf mass correspond to the mass of the model at the cooling curve, while hollow circles shows the mass of the hydrogen free core at the first thermal pulse. At the low mass end, the increase of the core mass during the TP-AGB is small, because of the small number of thermal pulses. However, as the initial mass increases, the number of thermal pulses also increases and so will the mass of the hydrogen free core.

\cite{romero15}  compare their results with those obtained by other authors. They found that, at the low mass end the initial-to-final mass agrees with that of \cite{weiss09}, especially for high metallicity progenitors, while for high masses and low metallicities the differences became more important. The white dwarf masses from \cite{weiss09} are smaller than those from \cite{romero15}, because \cite{weiss09} calculations allow overshooting during the TP-AGB, limiting the growth of the hydrogen-free core. In fact, the initial-to-final from \cite{weiss09} resembles the initial-to-final mass of \cite{romero15} for the core mass at the first thermal pulse. Note that pre-white dwarf ages for high mass progenitors, where the effects of overshooting during the TP-AGB are more important, do not change considerably with initial mass (see Tab. 1 from \citet{romero15}). 

\begin{figure}
\resizebox{\hsize}{!}{\includegraphics[clip=true]{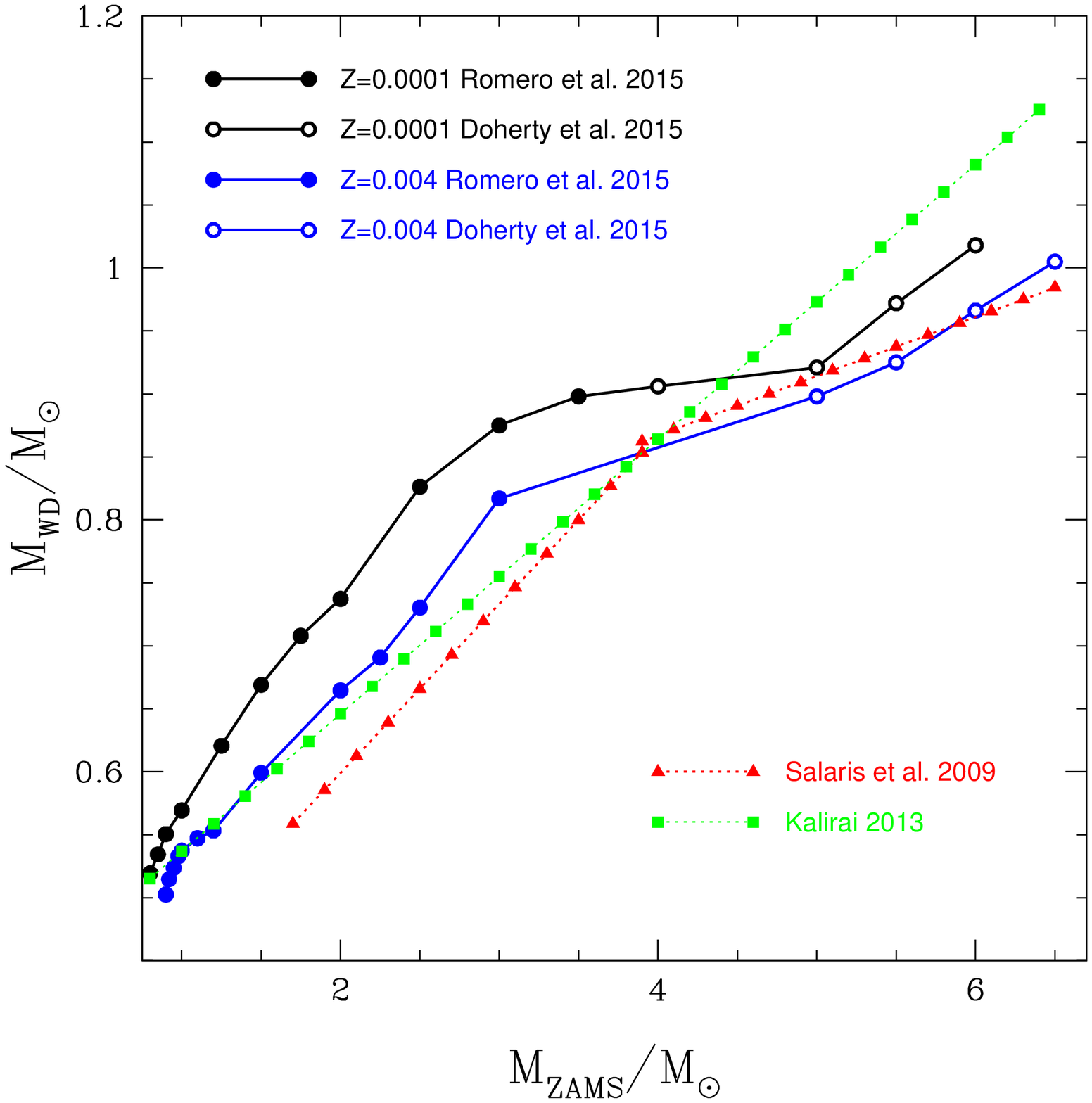}}
\caption{\protect
Initial - Final mass relation from \citet{romero15} (filled circles) and Doherty et al. (2015) (hollow circles) for the metallicities of $Z=0.0001$ and $Z=0.004$ and. The range of progenitor masses are 0.8 - 3.0\,$M_{\odot}$ for \citet{romero15} models and $\sim$4.0-7.0\,$M_{\odot}$ for \citet{doherty15} models. The red triangles and green squares represent the semi-empirical initial-to-final mass from \citet{salaris09} and \citet{kalirai13} that do not take different metallicities into account.
}
\label{massifkalsal}
\end{figure}

\cite{romero15} also compared sequences with initial mass 1$\,M_{\odot}$ and metallicity of Z=0.0001 and Z=0.02 using {\small MESA} \citep{paxton11,paxton13} including similar mass loss. As a result, they found that {\small LPCODE} lifetimes were consistent with those computed with {\small MESA}, resulting in differences smaller than $\sim0.1$\,Myr at $\sim$25\,000\,K on the white dwarf cooling curve. 

In order to extend our mass range, we included additional white dwarf sequences with stellar masses up to $\sim 1.0 M_{\odot}$, corresponding to progenitor masses up to $6M_{\odot}$ for $Z=0.0001$ and $7M_{\odot}$ for $Z=0.004$. We adopt the initial-to-final mass from \citet{doherty15} for carbon-oxygen white dwarf remnants with the metallicities of $Z=0.0001$ and $Z=0.004$. To compute the white dwarf evolution, we took a white dwarf model at high effective temperatures at the beginning of the cooling curve for the highest stellar mass evolutionary model available in our original grid, and artificially scaled the stellar mass from \citet{romero13}. We also changed the carbon/oxygen central composition to match that from \citet{doherty15} for each stellar mass and metallicity. The pre-WD ages were taken from \citet{doherty15} computations. Further evolution on the white dwarf cooling sequence was computed employing the LPCODE, including all physical ingredients considered in white dwarf computations from full evolution. Figure \ref{massifkalsal} shows the initial-to-final mass extended to progenitor masses of $\sim 7M_{\odot}$. Filled circles corresponds to full evolutionary computations (same as figure \ref{masstwd}) while hollow circles correspond to \citet{doherty15} results. We also compare with the semi-empirical initial-to-final mass from \citet{salaris09} and \citet{kalirai13}. Note that semi-empirical determinations of the initial-to-final mass relations do not take different metallicities into account.

We use the colours of the white dwarf cooling models consistent with those from recent results by \cite{tremblay11} that include the Lyman-$\alpha$ red wing calculations from \cite{kowalski06}. The $\rm HeH^{+}$ molecule in the equation of state, which becomes very important at $T_{\rm eff}<8\,000\,K$, is also included \citep{harris04, kilic10}. 

We also explored the possibility of white dwarf atmospheric models composed by a mixture of hydrogen and helium. The colours of the atmosphere models with hydrogen and helium mixed consider thick hydrogen layers ($q_H\!\equiv\!M_H/M_{\star}\!=\!10^{-4}$)  for low He/H ($\rm He/H\le1$). For higher He/H values, thin hydrogen layers ($q_H\!\equiv\!10^{-10}$) were considered. Those models were calculated ranging from $\rm log(He/H)=-2$ to $ \rm log(He/H)=8$.

\section{Data Analysis}
\label{analysis}

\cite{richer13} argued that the difficulty in age determination through the main sequence turn-off lies in the fact that the differences between two or more clusters can be due to metallicity and/or age differences. They compared the white dwarf cooling sequences of 47\,Tuc, NGC\,6397 and M\,4, considering the properties, distance modulus and reddening. They demonstrated that the white dwarf cooling sequences of those three clusters almost align perfectly in the colour-magnitude diagram, as seen in Fig. \ref{3clus}, even though they have different metallicities. This is consistent with the timescales of diffusion processes in the white dwarf atmospheres \citep{fontaine79}. \cite{richer13} claimed that white dwarf spectra and location on the colour-magnitude diagram do not depend on the cluster metal abundance, unless there is a possible dependence of the white dwarf mass on metallicity through the initial-to-final mass relation \citep{kalirai05,kalirai08,kalirai09}.  

\begin{figure}
\resizebox{\hsize}{!}{\includegraphics[clip=true]{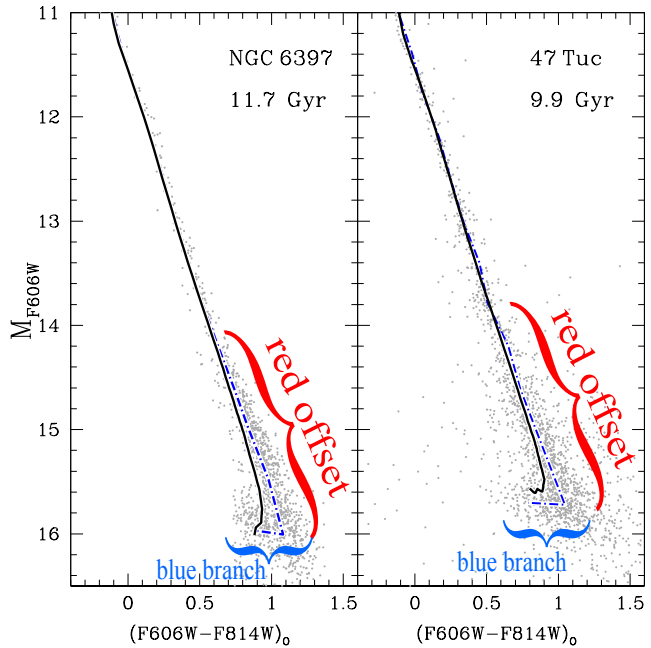}}
\caption{\protect
Our isochrone models (solid black line), with the ages determined by \citet{hansen13} over-plotted to the white dwarf cooling sequences of NGC6397 and 47Tuc. The blue dot-dashed line represents a mean ridge line of the data. The distance modulus and reddening correction are the ones by \citet{richer13}. For 47\,Tuc, the ``blue branch'' of the isochrone with 9.9\,Gyr model happens at a magnitude brighter than the one from the data. The red offset feature is also highlighted in the colour-magnitude diagram of both clusters.
}
\label{puremodels}
\end{figure}

Our first step was comparing our pure hydrogen isochrone models (Sect. \ref{models}), with the ages  determined by \cite{hansen13}, to the white dwarf cooling sequences of NGC\,6397 (11.7\,Gyr) and 47\,Tuc (9.9\,Gyr). In Fig. \ref{puremodels} we can see that, especially for 47\,Tuc, the ``blue branch'' of the model happens at a brighter magnitude than the one from the data. Also, if we compare the mean ridge line of the data (blue dot-dashed line) to the isocrhone models, for both clusters, Fig. \ref{puremodels} also highlights a change in the slope of the white dwarf cooling sequence after F606W$\sim$14.5, showing a trend for the models to be bluer (brighter) than the data, that is, a red offset of the data. 

Such trend could be caused by problems in the photometric calibration, so, before we began our analysis, we calculated the mean ridge line of the Small Magellanic Cloud (SMC) that is present in the non proper motion-corrected data of 47\,Tuc \citep[e.g.][]{kalirai12,richer13}, in an attempt to test this hypothesis. The mean ridge line of the SMC does not present the same trend towards redder colours as the one from the white dwarf cooling sequence of 47\,Tuc, indicating that poor photometric calibration is, apparently, not causing this slope change. 

This mismatch between models and data in the white dwarf cooling sequence is in the same line of problems found in the comparison of main sequence models to the data. For example, \cite{chen14} has shown that for low mass main sequence stars -- the coolest stars in the main sequence -- the data is redder than the models, even after they adopt better bolometric correction tables and new $T-\tau$ relations. They also mention that this problem seems to extend to other sets of models in the literature. 

Clearly, there are aspects of the construction of the models of white dwarf stars that need further investigation.  We estimate that $\sim19$\% of the total number of observed white dwarf stars lie on the red offset portion. One possible explanation for the red offset of the white dwarf 
stars could be a fraction of pure helium atmosphere white dwarf stars that 
would move the models to bluer colours. White dwarf masses would have to be $\sim$0.55\,M$_{\odot}$ at the bottom of the cooling sequence for the models to deviate as much as the data.

\citet{hansen04} and \citet{bedin09} included the fraction of helium atmosphere 
in their analysis of the white dwarf population of M4. \citet{hansen04} treated it 
as a free parameter and estimated an upper limit of 40\%, while \citet{bedin09} fixed 
the fraction at 30\%, the typical value for the DB/DA ratio in the disk population. However, 
\citet{davis09} analysed the spectra of 24 bright white dwarf stars in the line of sight 
of M4 and determined that all had hydrogen atmospheres. They argued that if all the 24 white 
dwarf stars were members of the cluster, the probability of observing helium atmosphere
white dwarf stars is $6\times 10^{-3}$. \citet{davis09} thoroughly discussed the probability
of finding helium atmosphere white dwarf stars in globular clusters, including results 
from other authors. They concluded that it is clear that the hydrogen/helium white dwarf 
stars atmosphere ratio in globular clusters is lower than in the field and that even though it is not impossible to form a non-hydrogen atmosphere in the cluster environment, the formation mechanism is clearly strongly suppressed. So, it is very unlikely that a fraction of 30 to 40\% of helium atmosphere white dwarf stars are observed in globular clusters.  

\citet{davis09} only used the spectra of hot white dwarf stars in globular clusters. 
However, there is little spectra information on white dwarf stars with effective temperatures 
lower than $\sim$ 5\,000\,K. \citet{limoges15} performed a census of white dwarf stars 
within 40\,pc of the Sun and found that probably most white dwarf stars with T$_{\rm eff} <$ 5\,000\,K have pure hydrogen atmospheres. However, they pointed out that the results below T$_{\rm eff}$ = 5\,000\,K should be considered with caution. Depending on the resolution, the spectra of ultra cool white dwarf stars do not show hydrogen or helium lines, thus providing no information on their surface 
gravities. So, inferences on the composition of the atmosphere can only be modelled with an
analysis of the energy distribution \citep{bergeron95}.  

\cite{koester76}, \cite{vauclair77} and \cite{dantona79} independently proposed the existence of a convective mixing between the massive helium layer and the thin hydrogen layer above it for cool white dwarf stars. When in the mixing process, the convection zone of a hydrogen atmosphere white dwarf star eventually reaches the underlying helium layer, helium will be brought to the surface through convective motion, resulting in the mixing of hydrogen and helium layers. \cite{bergeron90} showed that the surface gravities, inferred from spectroscopy of a sample containing 37 cool white dwarf stars ($T_{\rm eff}\la 12\,000\,K$), were significantly larger than the canonical value expected for these stars ($log\,\mathit{g}\sim8$), and that was interpreted as an evidence of the convective mixing between the hydrogen and helium layers. However, with high signal-to-noise, high-resolution spectroscopic observations for six cool DA white dwarf stars ($10\,720\,K\la\,T_{\rm{eff}}\la\,12\,630\,K$), obtained with the Keck I telescope, \cite{tremblay11} detected no helium in the spectra of any of the target stars, concluding that their helium abundance allowed to rule out the incomplete convective mixing scenario as the source of the high-log$\,\mathit{g}$ problem. Later, \citet{tremblay13, tremblay15} found that the the imprecise convection calculation was suppressing the convective mixing scenario. 

The results from \cite{tremblay11} do not rule out the hypothesis of  mixed atmosphere for ultra-cool white dwarf stars ($T_{\rm eff}\la\,5\,000\,K$).  Calculations by \cite{tremblay08} indicate that the effective temperature at which the mixing would occur depends on the thickness of the hydrogen envelope, meaning that, the thicker the envelope, the lower the mixing temperature. \cite{tremblay08} showed that the mixing should not occur if the mass of the hydrogen layer is larger than $M_H/M_{\star}\sim10^{-6}$. \cite{chen11} presented a theoretical analysis of white dwarf stars with the hydrogen layer mass between $10^{-7}$ and $10^{-11}M_{\star}$. They showed that a white dwarf star, upon convective mixing, always decreases the amount of hydrogen on the surface, but the spectral outcome and the change in the effective temperature depends on the mass of the hydrogen layer. 

The existence of white dwarfs with the thickness of the hydrogen layer between $10^{-9.5}< M_{\star}<10^{-4}$ was shown by \cite{castanheira08, castanheira09} and \cite{romero12,romero13}, through asterosismological studies. \cite{castanheira08} argue that this indicates that white dwarfs with an atmosphere of H, even if their total masses are close to the most probable value, may have formed with a H mass several orders of magnitude smaller than the value predicted by theory, i.e., it is likely that the mass loss during its evolution was actually more efficient than assumed by evolutionary models including mass loss of \cite{reimers77}, \cite{vassiliadis93} and \cite{groe09}. The most probable scenario is that the star might have experienced a late thermal pulse that consumed most of its hydrogen layer leaving a very thin ($10^{-8} -10^{-10}$) hydrogen layer. 

Preliminary tests with our models show that, for the mixing of hydrogen and helium to occur at
temperatures consistent with the change in slope of the models ($T_{\rm eff}\sim 5\,000\,K$), the mass of the hydrogen layer should be between $10^{-6}$ and $10^{-7}M_{\star}$, and the atmosphere would not be turned into pure helium. Instead, the atmosphere would be composed by hydrogen with traces of helium. When this mixing occurs, the atmosphere becomes more transparent and appears slightly hotter, causing a shift to a brighter magnitude; after that, the star continues to cool, but not as a pure hydrogen atmosphere anymore. This effect could be the reason of the observed slope change of the white dwarf cooling sequence from the globular clusters; however, this requires further investigation and proper modelling of the convective processes. 

Also, to perform an analysis of such cold white dwarf stars with mixed atmospheres, we should consider that, bellow 16\,000\,K, line broadening by neutral particles becomes important at helium rich atmospheres \citep{koester15} . However, only some He transitions have so-called self broadening theories, and they were tested for very low temperatures (300\,K). How calculations for appropriate temperatures would affect the models is still unknown.

Assuming that mixing of hydrogen and helium occurs in the atmosphere of the very cool white dwarf stars ($T_{\rm eff}\la\,6\,000\,K$), \citet{saumon14} studied the near-UV absorption in  white dwarf stars with models containing the revised calculations of $H_2-He$ collision-induced absorption opacity from \citet{abel12}. We compared our  single mass white dwarf model for 0.60$M_{\odot}$ with log(He/H)=0 to the one by \citet{saumon14} with the newest opacity, with the same parameters, and we found no difference between models with the new and old $\rm H_2-He$ collision induced absorptions for $T_{\rm eff}\ga$5\,000\,K. \citet{saumon14} mention the fact that $\rm H_2-H_2$ calculations are still in progress, so if they will help to solve the red offset problem is an important open question.

\subsection{Comparison of models with data}
\label{fit}

\citet{hansen07} pointed that an important part of modelling Monte Carlo simulations is the need to have the same observational scatter and incompleteness as the data. This is even more important for white dwarf stars, as even the brightest ones could be lost if they are close to a bright main sequence star. To take that into account, \citet{hansen07,hansen13} performed artificial star tests so they could  measure the recovery fraction and determine the association between input and observed magnitude. 

To obtain the cluster parameters we used our isochrone models to generate 
Monte Carlo realisations of the white dwarf populations and compared them 
with the observed cooling sequences. 

Our Monte Carlo simulations follow an inverse transform sampling \citep{fisherman96}. For the mass interval represented in the observed luminosity function, we are assuming a single burst star formation and a \citet{salpeter55} distribution (dN/dM$\propto M^{-\alpha}$) for the present day local mass function. We allowed $\alpha$ to vary from 1.6 to 4.0. As we are not considering any dynamical effects in our models, we are only obtaining the present day local mass function. As the mass of the progenitor star represented in the observed luminosity function from the white dwarf stars is only from 0.85 to $\sim$6.00, we assumed a single burst star formation. We also adopt the metallicity dependent initial-to-final mass relation from \cite{romero15} and generate models with a range of distance modulus and reddening appropriate to each cluster.

To account for the size of the photometric errors, for each model white dwarf star we 
add a photometric uncertainty that was determined through artificial star tests consistent 
with NGC\,6397 and 47\,Tuc \citep{hansen07,hansen13}. The artificial star tests provide not 
only the degree of correlation between the input and observed magnitudes, but also the
recovery fraction, i.e., the photometric completeness of the data, which is also taken 
into account in our simulations. This process was applied for both F606W and F814W filters.
Thus, each realisation includes realistic photometric scatter and the correct 
level of completeness. Our Monte Carlo simulations 
do not account for the effect of multiple star formation bursts, unresolved binaries or multiple populations.

\begin{figure}
\resizebox{\hsize}{!}{\includegraphics[clip=true]{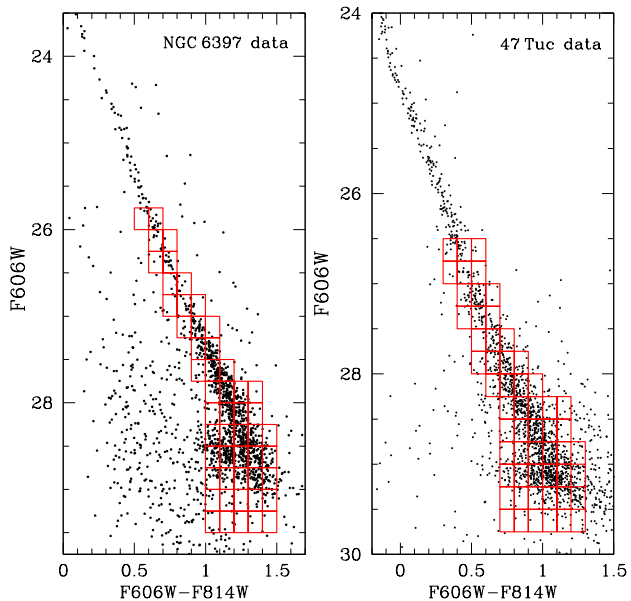}}
\caption{\protect
The red grid overlaid on the white dwarf cooling sequences of NGC\,6397 and 47\,Tuc illustrating the way the data were binned in order to compare the observations to the Monte Carlo simulations with Hess diagrams. Each bin measures 0.25\,mag in magnitude and 0.1\,mag in colour.
}
\label{grid}
\end{figure}

To reduce model uncertainties we generated simulations with 20\,000 stars. Observed and model
luminosity functions are built with bins of 0.20\,mag; Hess diagrams are binned with a grid 
width of 0.1 mag in F606W-F814W and 0.25\,mag in F606W, a grid similar to that used in previous 
analysis of the same data \citep{hansen07,hansen13,garcia-berro14,torres15}, as shown in Fig. \ref{grid}. Simulations are normalised to have the same number of stars as the
corresponding cluster.

To compare the Hess diagrams of the models and the observations we used the reduced ${\chi^2_{red}}$ which is the sum of the residuals taking the density of the observed and model of each grid cell into account, according to:
\begin{equation}
{\chi^2}=\sum_{i}\frac{(O_i-E_i)^{2}}{\sigma^ 2_{O_i}}
\end{equation}
where $E_i$ is the number of events expected according to the model  $O_i$ is the number observed in the ith bin and, as we assume our data follow a Poisson distribution, $\sigma_{O_i}$ is the uncertainty in the number of observed stars ($\sqrt{O_i}$).

We compute independent $\chi^2$ tests for the luminosity function and for the Hess diagram. The $\chi^2$ tests for the colour functions is not computed because the red offset masks the information. After that we obtained the reduced  $\chi^2$  for both approaches. Hereafter, we followed the  process adopted by \citet{garcia-berro14} and \citet{torres15}, and the reduced $\chi^2$ values were normalised to the minimum value for each of the tests and added quadratically. \citet{torres15} argued that  although they have used a $\chi^2$ test, their final aim was to estimate the values of the free parameters that best fit the observed data, rather than obtaining an absolute probability of agreement of  the models with the observed data. That is precisely the same goal we have in our analysis, therefore using the same kind of analysis employed by \cite{garcia-berro14} and \citet{torres15} is a meaningful approach.

To begin our analysis we obtained the distance modulus and reddening from the data by comparing the Monte Carlo simulations to the data of the white dwarf stars hotter than $\sim$5\,500\,K, i.e., the top of the white dwarf cooling sequence. Also, we compare our age and $\alpha$ determinations to the ones found if we fix the distance modulus and the reddening obtained through isochrone fittings  to the main sequence for NGC\,6397 ($\mu_{0}$=12.07$\pm$0.06 and A$_V$=0.56$\pm$0.06, \citealt{richer08}) and 47\,Tuc ( $\mu_{0}$=13.26 and A$_V$=0.07, \citealt{dotter10}).

The results from our analysis for NGC\, 6397 and 47\,Tuc, performed as explained above, are shown in Sect.  \ref{ngc6397section} and \ref{47tucsection}.

\subsubsection{NGC\,6397}
\label{ngc6397section}

The fitting of the white dwarf stars hotter than 5\,000\,K to our Monte Carlo simulations resulted in a distance modulus of $\mu_{0}$=11.85$\pm$0.04 and A$_V$=0.64$\pm$0.04, or E(B-V)=0.21 (assuming $R_{V}=3.1$). These values are in agreement, within the uncertainties, with the ones presented by \citet{richer13} and references therein. 

\begin{figure}
\resizebox{\hsize}{!}{\includegraphics[clip=true]{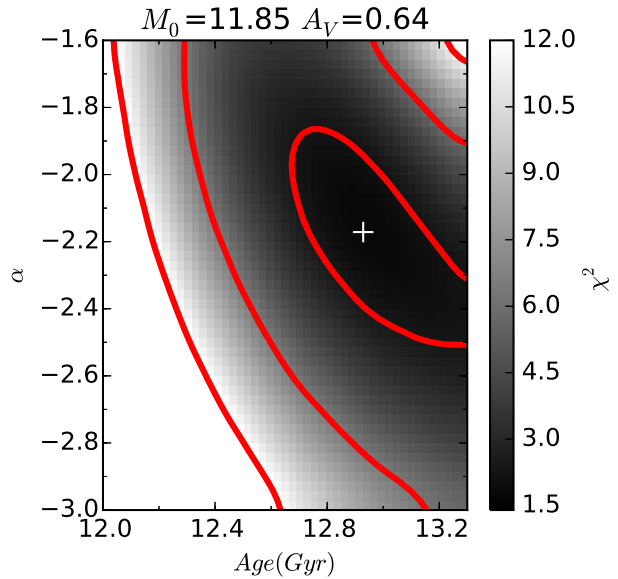}}
\caption{\protect
$\chi^2$ (grey scale) for age and $\alpha$ for NGC\,6397 with our determination of distance modulus and reddening considering only the white dwarf stars hotter than 5\,000\,K. The red lines represent the regions with 68\%, 95\% and 99\% confidence level and the white cross represents the best fit model.
}
\label{ourmo6397}
\end{figure}

After that we performed the fitting of our Monte Carlo simulations to the data allowing age and $\alpha$ to vary and keeping the distance modulus and reddening as fixed parameters. The range of ages for NGC\,6397 was set to vary from 12.0 to 13.3\,Gyrs and $\alpha$, of the present day local mass function, could vary from 3.0 to 1.6. The $\chi^2$  is presented in Fig. \ref{ourmo6397}, we also show the curves enclosing the regions with 68\%, 95\% and 99\% confidence level using lines. The model that best fit the data has an age of 12.93$_{-0.21}^{+0.37}$\,Gyr and $\alpha$=2.17$_{0.30}^{0.34}$ is marked as a white cross. 

\begin{figure}
\resizebox{\hsize}{!}{\includegraphics[clip=true]{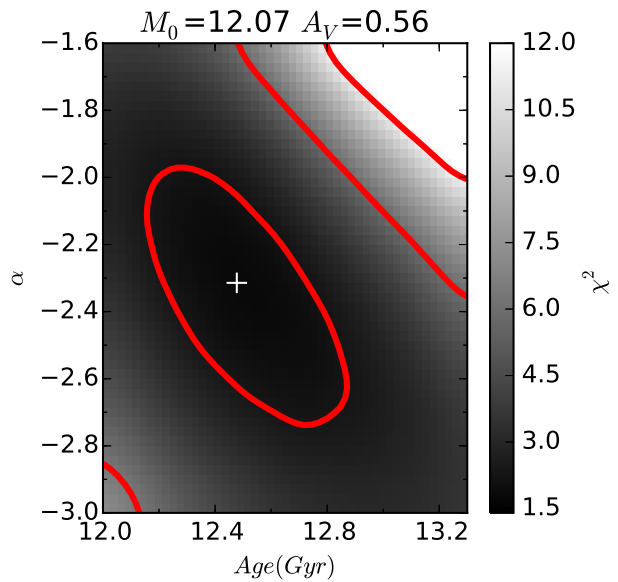}}
\caption{\protect
$\chi^2$ (grey scale) for age and $\alpha$ for NGC\,6397 with the determination of distance modulus and reddening obtained by \citet{richer08} through the main sequence fitting. The red lines represent the regions with 68\%, 95\% and 99\% confidence level and the white cross represents the best fit model.
}
\label{richer08mo}
\end{figure}

We also tested fixing the distance modulus and the reddening to the ones by \citet{richer08}, obtained by fitting models to the main sequence of NGC\,6397 data obtained with Hubble Space Telescope, $\mu_{0}$=12.07$\pm$0.06 and A$_V$=0.56$\pm$0.06. In that way we test the effect of keeping the coherence between the main sequence and the white dwarf cooling sequence. The results for the $\chi^2$ is presented in Fig. \ref{richer08mo} . The model that we found as the one that best fitted the data in this case has an age of 12.48$_{-0.26}^{+0.34}$\,Gyr and $\alpha$=2.31$_{-0.33}^{+0.39}$.

When we compared the age and $\alpha$ obtained with our distance modulus and reddening to the results obtained with the distance modulus and reddening consistent to the main sequence by \citet{richer08} we obtain differences of 0.45\,Gyrs in age and 0.14 in $\alpha$. Both determinations are consistent, within the uncertainties, and the age difference is smaller than the uncertainties of our models of the order of  0.50\,Gyrs.

\begin{figure}
\resizebox{\hsize}{!}{\includegraphics[clip=true]{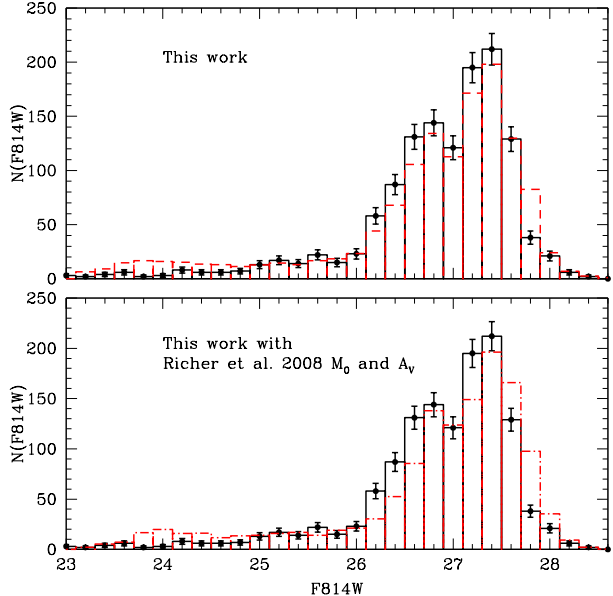}}
\caption{\protect
Luminosity function of NGC\,6397 (solid black line) compared to the best fit models. At the top panel we show the model with  12.93$_{-0.21}^{+0.37}$\,Gyr and $\alpha$=2.17$_{-0.30}^{+0.34}$, obtained with our distance modulus and reddening determination (dashed red line). While in the lower panel we show the luminosity function of our model with the distance modulus and reddening obtained by \citet{richer08} with  12.48$_{-0.26}^{+0.34}$\,Gyr and $\alpha$=2.31$_{-0.33}^{+0.39}$ (dot-dashed red line). The error bars are Poisson errors. 
}
\label{lfourricher08}
\end{figure}

Figure \ref{lfourricher08} shows the luminosity function of our best fit Monte Carlo simulations compared to the data. As our simulations have 20\,000 stars, we normalised the simulations to the total area of the data.  At the top panel we show the best fit model obtained with our distance modulus and reddening determination. In the lower panel we show the luminosity function of our model with the distance modulus and reddening obtained by \citet{richer08}. 

It is clear, from Fig. \ref{lfourricher08}, that the model with our distance modulus and reddening determinations presents a better fit to the data, mainly for the fainter white dwarf stars. However, for the brightest white dwarf stars, both models present a number of stars that is higher than the one observed in the data. 

\begin{figure}
\resizebox{\hsize}{!}{\includegraphics[clip=true]{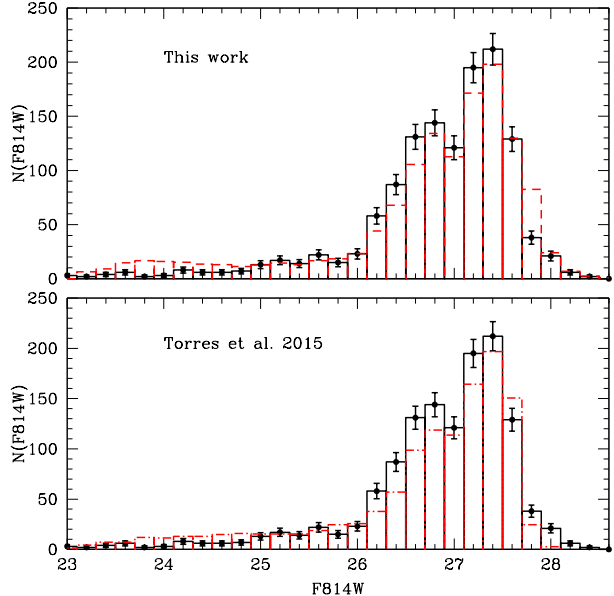}}
\caption{\protect
 In the top panel, the normalised luminosity function of our best fit model (dashed red line) compared to the data of NGC\,6397 (solid black line). In the lower panel we show the luminosity function presented in Fig. 1 of \citet{torres15} compared to the same data (dot-dashed red line). The error bars are Poisson errors.
}
\label{lfourtorres15}
\end{figure}

\begin{figure}
\resizebox{\hsize}{!}{\includegraphics[clip=true]{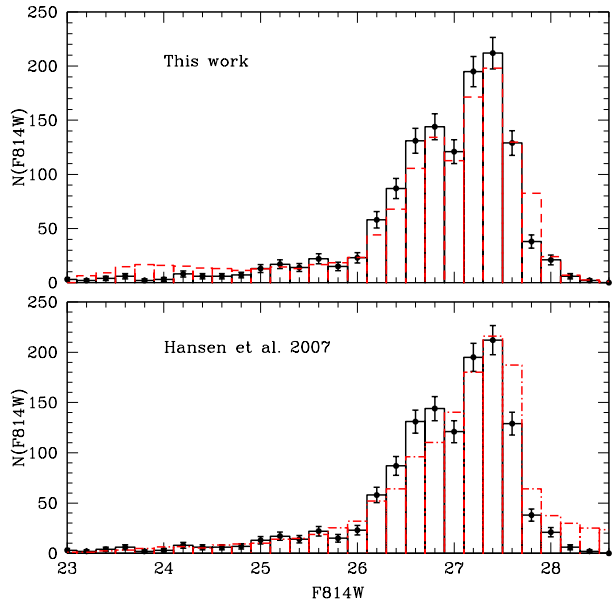}}
\caption{\protect
In the top panel, the normalised luminosity function of our best fit model (dashed red line) compared to the data of NGC\,6397 (solid black line). In the lower panel we show the luminosity function presented in Fig. 15 of \citet{hansen07} compared to the same data (dot-dashed red line). The error bars are Poisson errors.
}
\label{lfourhansen}
\end{figure}

We also compared our simulated luminosity function to the ones from \cite{torres15} and \citet{hansen07}. In figs. \ref{lfourtorres15} and \ref{lfourhansen} we show, in the top panel, the luminosity function of our best fit model compared to the data. In the lower panel of Fig. \ref{lfourtorres15} we show the luminosity function of the best fit model by \citet{torres15}  presented in their  Fig. 1, compared to the data, while in  the lower panel of Fig. \ref{lfourhansen} we show the luminosity function of the best fit model by \citet{hansen07}  presented in their  Fig. 15.   Both \citet{torres15} and \citet{hansen07} simulations present  a better fit for the hot white dwarf stars than our models with more free parameters. However, the crystallisation peak at F814W$\sim$26.8  is not present in \citet{torres15} and \citet{torres15} simulations. Our simulation clearly presents a peak at the same magnitude as the crystallisation peak, demonstrating the quality of our models.

\begin{figure}
\resizebox{\hsize}{!}{\includegraphics[clip=true]{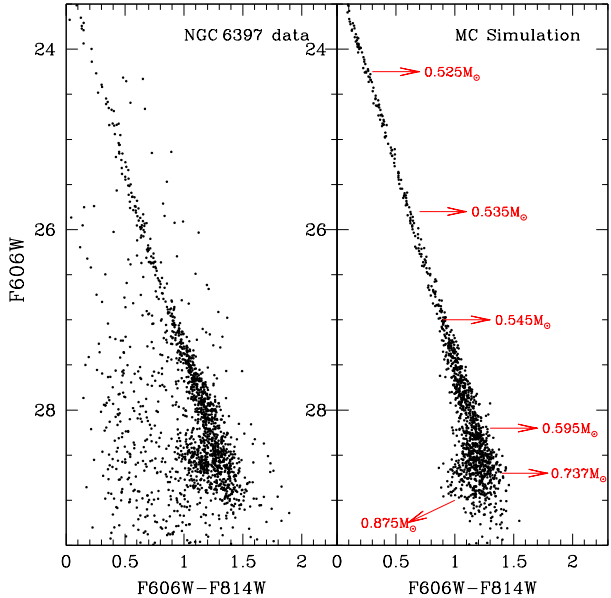}}
\caption{\protect
The left panel shows the observed white dwarf cooling sequence of NGC\,6397, after the star-galaxy separation described in \citet{hansen07}. The the right panel shows the Monte Carlo simulation with 12.93\,Gyr, that best fitted the data with our distance modulus and reddening determination. The red labels illustrate the spread in mass, according to the models, ranging from 0.525\,M$_{\odot}$, at the top of the cooling sequence, which is consistent with spectroscopic results, to 0.875\,M$_{\odot}$ at the bottom.
}
\label{ngc6397mc}
\end{figure}

The comparison between the white dwarf cooling data and the Monte Carlo model that best fits the data presents an excellent agreement, as can be seen in Fig. \ref{ngc6397mc}, for NGC\,6397. We also illustrate the spread in mass (red labels), according to the models, ranging from $\sim$0.525\,M$_{\odot}$, at the top of the cooling sequence, which is consistent with spectroscopic results, to $\sim$0.875\,M$_{\odot}$ at the bottom. 

The turn to the blue is better reproduced in our models, if we compare to the one presented by \citet{hansen07}, in their Fig.\,11, where the models did not fully reproduced the clump of stars blueward of the faint end of cooling sequence and the most massive white dwarf star in their model was $\sim$0.62\,M$_{\odot}$. When compared to the models by \citet{torres15}, our colour magnitude diagram is just as good but, as discussed previously, our luminosity functions present features consistent with the data that are not apparent in \citet{torres15} simulations.

\subsubsection{47\,Tuc}
\label{47tucsection}

By performing the fitting of the white dwarf stars hotter than 5\,000\,K of 47\,Tuc to our models we obtained the distance modulus of $\mu_{0}$=13.28$_{0.03}^{0.06}$ and A$_V$=0.14$\pm$0.01, or E(B-V)=0.045. These values are in agreement with the ones presented by \citet{richer13} and references therein. 

\begin{figure}
\resizebox{\hsize}{!}{\includegraphics[clip=true]{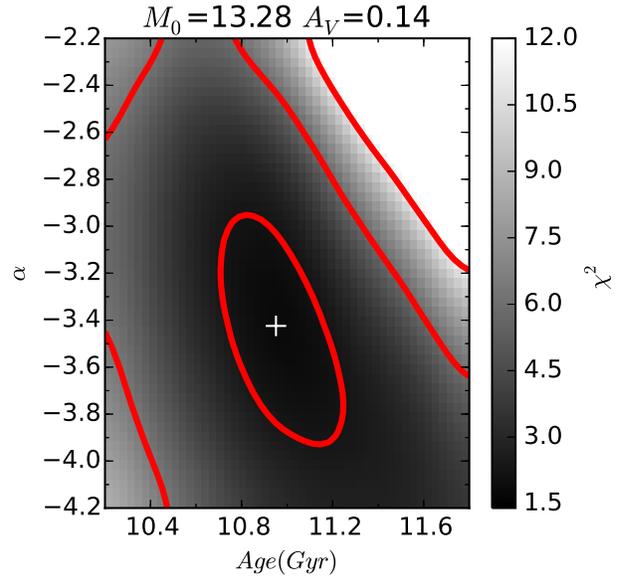}}
\caption{\protect
$\chi^2$ (grey scale) for age and $\alpha$ for 47\,Tuc with our determination of distance modulus and reddening considering only the white dwarf stars hotter than 5\,000\,K. The red lines represent the regions with 68\%, 95\% and 99\% confidence level and the white cross represents the best fit model.
}
\label{ourmo47}
\end{figure}

So we performed the fitting of our Monte Carlo simulations to the data allowing age and $\alpha$ to vary and keeping the distance modulus and reddening as fixed parameters. The range of ages for 47\,Tuc was set to vary from 10.2 to 12.8\,Gyrs and $\alpha$, of the present day local mass function, could vary from 4.2 to 1.6. The $\chi^2$ is presented in Fig. \ref{ourmo47}. The model that best fit the data has an age of 10.95$_{-0.15}^{+0.21}$\,Gyr and $\alpha$=3.42$_{-0.46}^{+0.50}$ is marked as a white cross. 

\begin{figure}
\resizebox{\hsize}{!}{\includegraphics[clip=true]{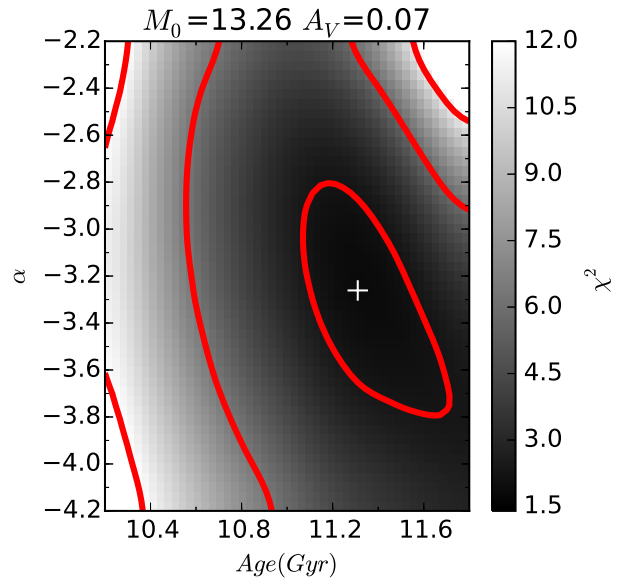}}
\caption{\protect
$\chi^2$ (grey scale) for age and $\alpha$ for 47\,Tuc with the determination of distance modulus and reddening obtained by \citet{dotter10} through the main sequence fitting. The red lines represent the regions with 68\%, 95\% and 99\% confidence level and the white cross represents the best fit model.
}
\label{dotter10mo}
\end{figure}

Again, to keep the coherence between the main sequence and the white dwarf cooling sequence, we fixed the distance modulus and the reddening to the ones by \citet{dotter10}, obtained by fitting models to the main sequence of 47\,Tuc data obtained with Hubble Space Telescope, $\mu_{0}$=13.26 and A$_V$=0.07. The results for the $\chi^2$ is presented in Fig. \ref{dotter10mo} . The model that we found as the one that best fitted the data in this case has an age of 11.31$_{-0.17}^{+0.36}$\,Gyr and $\alpha$=3.26$_{-0.44}^{+0.53}$.

Comparing the results obtained with our distance modulus and reddening to the results obtained with the distance modulus and reddening consistent to the main sequence by \citet{dotter10} we notice difference of 0.36\,Gyrs in age and 0.16 in $\alpha$. Such differences in age are smaller than the uncertainties of our models of the order of 0.50\,Gyrs. 

\begin{figure}
\resizebox{\hsize}{!}{\includegraphics[clip=true]{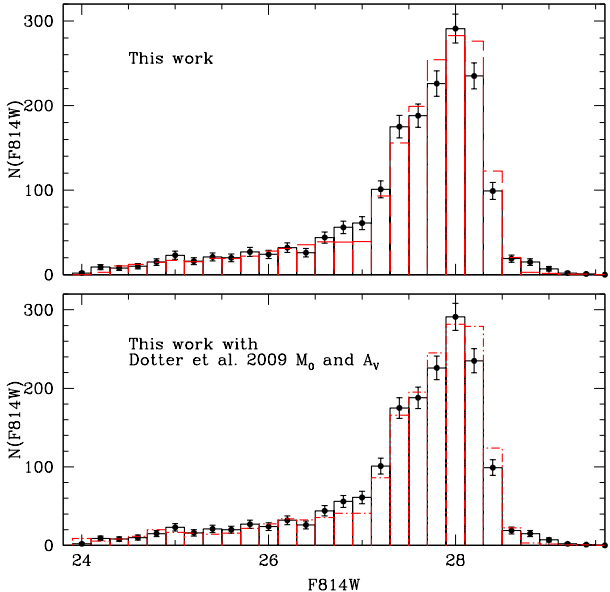}}
\caption{\protect
The normalised luminosity function of 47\,Tuc (solid black lines) compared to the best fit models. At the top panel we show the model with  10.95$_{-0.15}^{+0.24}$\,Gyr and $\alpha$=3.42$_{-0.46}^{+0.50}$, obtained with our distance modulus and reddening determination (dashed red lines). While in the lower panel we show the luminosity function of our model with the distance modulus and reddening obtained by \citet{dotter10} with 11.31$_{-0.36}^{+0.17}$\,Gyr and $\alpha$=3.26$_{-0.44}^{+0.53}$ (dot-dashed red lines). The error bars are Poisson errors. 
}
\label{lfourdotter47}
\end{figure}

\begin{figure}
\resizebox{\hsize}{!}{\includegraphics[clip=true]{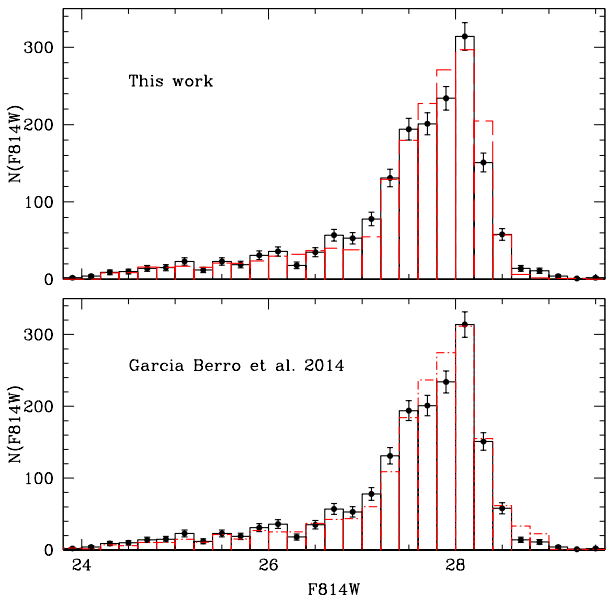}}
\caption{\protect
In the top panel, the normalised luminosity function of our best fit model (dashed red lines) compared to the data of 47\,Tuc (solid black lines). In the lower panel we show the luminosity function presented in Fig. 4 of \citet{garcia-berro14}, normalised to the total number of stars in our data (dot-dashed red lines), compared to the same data. The error bars are Poisson errors.
}
\label{lfourgarcia47}
\end{figure}

As \citet{hansen13} did not show their luminosity function and colour-magnitude diagrams of their simulations of 47\,Tuc,  we could only compare our simulated luminosity function to the ones from \cite{garcia-berro14}. In Fig. \ref{lfourgarcia47} we show, in the top panel, the luminosity function of our best fit model compared to the data. In the lower panel of Fig. \ref{lfourgarcia47} the luminosity function of the best fit model by \citet{garcia-berro14} presented in their  Fig. 4, compared to the data. As \citet{garcia-berro14} did not use the proper motion correction, their luminosity function had a total area slightly larger than the one from our sample. So we normalised their luminosity function to the total number in our data. It is clear that our models present a very similar agreement to the data when compared to the models from \citet{garcia-berro14}. One feature that must be emphasised is that, even though the effect is present in our models, neither the models nor the data of 47\,Tuc present a clear crystallisation peak.

 \begin{figure}
\resizebox{\hsize}{!}{\includegraphics[clip=true]{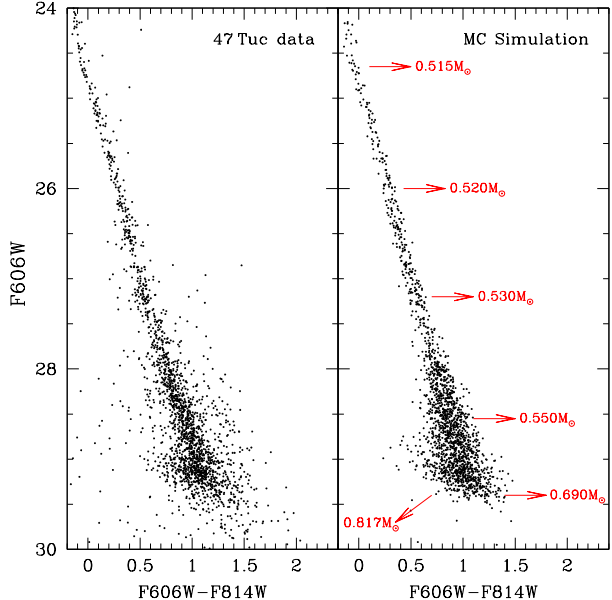}}
\caption{\protect
The left panel shows the observed white dwarf cooling sequence of 47\,Tuc with the proper motion selected stars. The right panel shows the Monte Carlo simulation with 11.20\,Gyr, that best fitted the data.  The red labels illustrate the spread in mass, according to the models, ranging from 0.515\,M$_{\odot}$, at the top of the cooling sequence, which is consistent with spectroscopic results, to 0.817\,M$_{\odot}$ at the bottom.
}
\label{47tucmc}
\end{figure}

When we compare  the white dwarf cooling data and the Monte Carlo simulation that best fits the data we find  an excellent agreement, as can be seen in Fig. \ref{47tucmc}, for 47\,Tuc. The spread in mass (red labels), according to the models, is ranging from $\sim$0.515\,M$_{\odot}$, at the top of the cooling sequence, which is consistent with spectroscopic results, to $\sim$0.817\,M$_{\odot}$ at the bottom.When compared to the models by \citet{garcia-berro14}, our colour magnitude diagram presents and excellent agreement, also for the hotter white dwarf stars. We could not compare our luminosity function to the ones obtained by \citet{hansen13} because they do not show this important result in their paper. Also, as the colour-magnitude diagram of the simulations from \citet{hansen13} is not shown, we can not see if the observed blue turn feature is  present in their models for 47\,Tuc. 

\subsubsection{M\,4}
\label{m4section}

With only 14 HST orbits obtained for M\,4, \cite{bedin09} were able to reach the blue turn of the white dwarf cooling sequence. The data obtained in filters F606W and F775W has poorer quality and much less stars, if compared with the data from NGC\,6397 and 47\,Tuc. The small number of stars along with the large scatter presented in the white dwarf cooling sequence prevents us from performing an analysis similar to the one we did for NGC\,6397 and 47\,Tuc. The chi-square analysis requires a minimum number of points to be reliable \citep{press07}, and that is not the case for this data. 

\begin{figure}
\resizebox{\hsize}{!}{\includegraphics[clip=true]{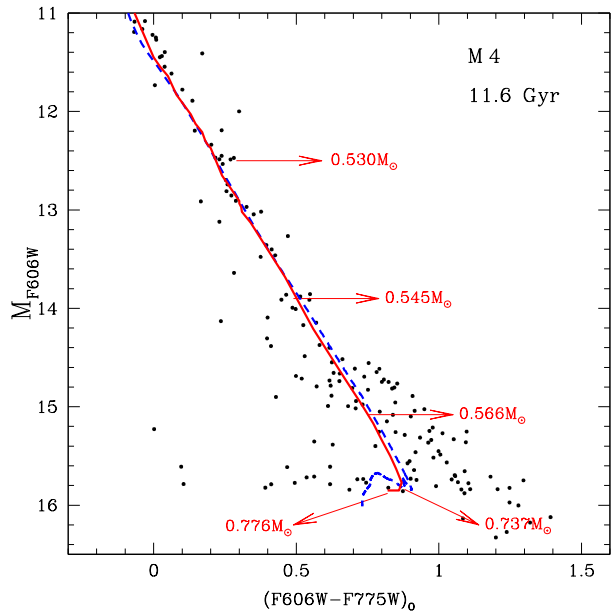}}
\caption{\protect 
Our isochrone model (solid red line) over-plotted with the observed white dwarf cooling sequence of M\,4 using the age, distance modulus and reddening correction obtained by \citet{richer13}. Also shown is the white dwarf isochrone obtained by \citet{salaris10} (dashed blue line), which points out the consistency between both models.    
}
\label{m4}
\end{figure}

Nevertheless, in order to compare our models to the data, in Fig.\,\ref{m4} we over-plotted the data using our isochrone model with 11.60\,Gyr (solid red line), the age determined by \citet{bedin09}, with the distance modulus and reddening from \citet{richer13}. It is noticeable that the model reproduces the blue turn very well, even though the blue branch of our model is shorter than the data because the photometric scatter for these data is very high and currently the highest progenitor mass we have in our model, for the metallicity of M\,4, is 2.25\,$M_{\odot}$. This model shows that the range of masses of the white dwarf stars is between $\sim 0.527M_{\odot}$ to $\sim 0.776M_{\odot}$. Again, the mass at the top of the white dwarf cooling sequence is consistent with the results from spectroscopy \citep{moehler04,kalirai09}.

We also over-plotted a 11.6\,Gyr white dwarf isochrone obtained by \citet{salaris10} (dashed blue line) to compare with our model. Although \citet{salaris10} models have a wider blue branch, because the highest progenitor mass in their model is $\sim6\,M_{\odot}$, the shape of their model is completely consistent with ours and, also, \citet{salaris10} models show that the blue turn is due the increasing masses.

\section{Discussion}
\label{discussion}

Our age determinations are consistent with those obtained by 
\citet{vandenberg13} using what they call the improved version 
of the $\Delta V^{HB}_{TO}$ method. They found13.00$\pm$0.25\,Gyr for NGC\,6397 and 11.75$\pm$0.25\,Gyr for 47\,Tuc that, within the uncertainties, agree with 
our values. In addition, our age value for 47\,Tuc is consistent, within 
the uncertainties, with the $12.0\pm0.5$\,Gyr determined by \citet{garcia-berro14} also using the cooling sequence method. And our determination
for NGC\,6397 also agrees with the 12.8$^{+0.50}_{-0.75}$\,Gyr determined by \citet{torres15} using the white dwarf cooling sequence.

\begin{figure}
\resizebox{\hsize}{!}{\includegraphics[clip=true]{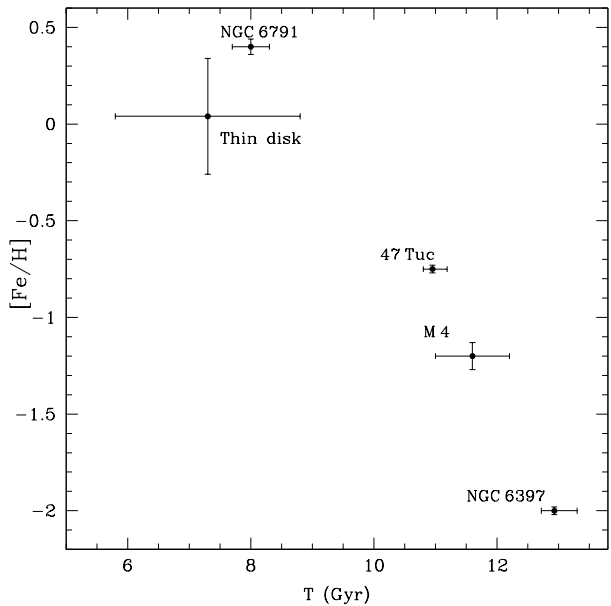}}
\caption{\protect
Age-metallicity relation based on ages determined with white dwarf cooling sequence and the metallicity for Galactic population determined from main sequence stars. The ages of NGC\,6397 and 47\,Tuc are the ones determined in the present work. The ages of NGC\,6791 and the thin disk were taken from \citet{hansen13} and references therein. For M\,4 the age is the one determined by \citet{bedin09}.
}
\label{agemet}
\end{figure}

To demonstrate the effect of our age determinations on the age-metallicity relation, we show in Fig. \ref{agemet} the age, determined through white dwarf cooling sequence, versus the metallicity for Galactic population, similar to what was done by \cite{hansen13}. It is possible to see that M\,4 has an intermediate metallicity when compared to 47\,Tuc and NGC\,6397, and, apparently, also an intermediate age. This reinforces the age-metallicity relation stating that the metal-poor formed earlier than the metal-rich globular clusters.

The comparison between the white dwarf cooling data and the Monte Carlo model that best fits the data presents an excellent agreement, as can be seen in Figs. \ref{lfourricher08} and \ref{ngc6397mc}, for NGC\,6397, and Figs. \ref{lfourdotter47} and \ref{47tucmc}, for 47\,Tuc. The turn to the blue is better reproduced for NGC\,6397, if we compare to the one presented by \citet{hansen07}, their Fig.\,11, where their models did not fully reproduced the clump of stars blueward of the faint end of cooling sequence. However, the scattering in the model of 47\,Tuc seems to be higher than the one from the data, indicating that the scatter of the artificial star tests could be overestimated. Also, the red offset problem is not corrected even when we include the photometric scatter in the models. 

Our Monte Carlo Simulations do not include multiple star formation bursts, unresolved binaries or multiple populations. Thus, the small discrepancies between our models and the data may be related to the lack of those effects.We discuss more in the following paragraphs.

\citet{torres15} found a low value ($\sim4\%$) for the binary fraction in NGC\,6397, which 
is very similar to the $\sim4.5\%$ fraction found for the main sequence stars by \citet{ji15}. The latter authors also determined the binary fraction for 47\,Tuc as being $\sim 3\%$,
arguing that, for globular clusters, the binary fraction slowly decreases with dynamical 
age, i.e., above 10 relaxation times, there should be no clusters with binary fractions 
above 6\%.

\cite{garcia-berro14} find that a burst of star formation for 47\,Tuc occurred between 0.7\,$\pm$0.5Gyr and 1.0$\pm$0.5\,Gyr, depending on the distribution employed. \cite{torres15} also attempted to determine the burst of star formation for NGC\,6397. However, their best fit obtained the duration of the burst of star formation as being $\Delta t$=3.3\,Gyr. They argued that, with the high accuracy photometric data obtained with Hubble Space Telescope, such an extended episode of star formation would be detectable in the colour-magnitude diagram of main sequence stars, which is not the case. So, they conclude that the photometric data available for the white dwarf stars of NGC\,6397 is not appropriate for this kind of analysis and they adopt a initial burst of star formation as being 1.0\,Gyr. 

\begin{figure}
\resizebox{\hsize}{!}{\includegraphics[clip=true]{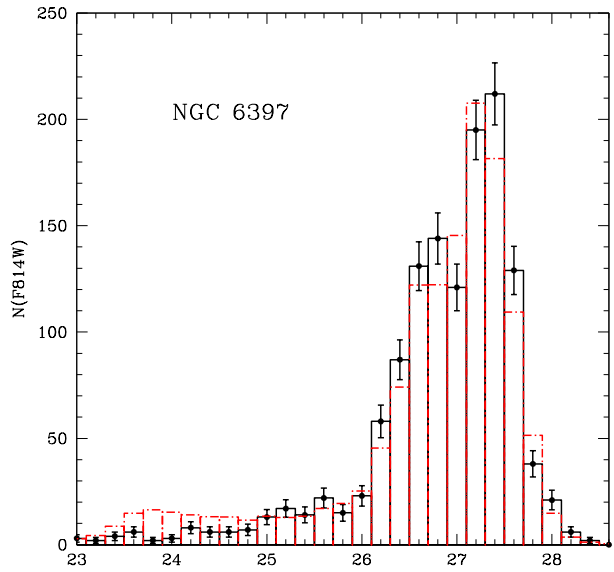}}
\caption{\protect
Comparison between the observed luminosity function from NGC\,6397 (solid black line) to a simulated one (dashed red line) with two bursts of star formation with 1\,Gyr difference (11.93\,Gyr and 12.93\,Gyr). The crystallization peak is not present in the model, indicating that the age difference between two bursts should be smaller than 1\,Gyr for NGC\,6397.
}
\label{burst}
\end{figure}

In Fig. \ref{burst} we show a comparison between the observed luminosity function from NGC\,6397 to a simulated one with two bursts of star formation with 1\,Gyr difference (11.93\,Gyr and 12.93\,Gyr). We notice that this model does not present a feature that is clearly present in the data: the crystallization peak. This feature is present in our single burst star formation model presented in Fig. \ref{lfourricher08}. That could be an indication that the age difference between two bursts, if present, should be smaller than 1\,Gyr for NGC\,6397. 

\cite{milone12} found that for certain filters, the photometry of 47\,Tuc splits not only the main sequence in two branches, but also the sub-giant, red-giant regions and on the horizontal branch. Their results indicated that the differences between the two main sequences could be explained in terms of two stellar generations, one of them being formed by the primordial He, and O-rich/N-poor stars, and the other corresponding to a second generation population that was enriched in He and N but depleted in O. The He content would be Y=0.256 for the primordial population and Y=0.272 for the second generation. \cite{hansen13} accounted for He enrichment consistent with the one obtained by \cite{milone12}, but they concluded that the resulting differences in the models do not lead to significant changes in the cooling age. 

When we determine the distance modulus and reddening with the white dwarf stars hotter than 5\,500\,K, our reddening determinations are systematically smaller than the ones determined with the main sequence fitting, with a $\Delta_{colour}\sim$0.07. This could indicate that the white dwarf stars hotter than 5\,500\,K are somehow slightly redder than the main sequence stars.

For NGC\,6397 and 47\,Tuc clusters our Monte Carlo simulations present a strong turn to the blue at the faint magnitudes, caused by increasing masses, and the temperature at this point is $\sim$4\,300\,K. The collision-induced absorption, as it is currently described \citep[e.g.][]{borysow89,borysow00}, would only be noticeable in infra-red colours ($M_K, J-K$), taking place at much lower temperatures, causing the second turn, but now to red colours as argued by \cite{bono13}. They claim that, at fainter magnitudes ($M_{\rm Bol}\sim16$), the white dwarf isochrone becomes populated by more massive white dwarf stars, and those are less blue because of the different cooling speed and onset of the collision-induced absorption, causing a red turn in the white dwarf cooling sequence for near infra-red isochrone models. 

When we focus on the slope of the present day local mass function, we find that $\alpha$=2.17$_{-0.30}^{+0.34}$ for NGC\,6397, that is slightly different from the canonical $\alpha$=2.35 from \citet{salpeter55}, and consistent with a top-heavy type, as has already been suggested by \citet{richer08} and references therein. In contrast, the value of $\alpha$=3.42$_{-0.46}^{+0.50}$ obtained as the best fit to 47\,Tuc data is very different from the canonical. However, it must be noted that we are analysing data from a single region of the cluster. The very high alpha determined for 47\,Tuc indicates a bottom heavy present day local mass function in that portion of the cluster, i.e., the present day local mass function of 47\,Tuc is
deficient in more massive white dwarf stars. Such a lack of massive white dwarf stars 
could be related to the diffusion due to gravitational relaxation that was detected for 
47\,Tuc by \citet{heyl15}. By analysing the data from the core of 47\,Tuc, \citet{heyl15} 
detected that the spatial distribution of young (less massive) white dwarf stars
is significantly more centrally concentrated than the older (more massive) ones, 
indicating that the white dwarf distribution seems to be more radially diffuse with 
increasing age \citep{heyl15}. As the data we are using in our analysis is very close to 
the centre (r$\sim$=8.8\,pc), while the tidal radius is r$_t\sim$52.0\,pc, a bottom heavy present 
day local mass function is another indicator of the lack of massive white dwarf  stars near the centre of 47\,Tuc.

A feature highlighted in Fig. \ref{ngc6397mc} and Fig. \ref{47tucmc} is that the models, 
built with the initial-to-final mass relation of \cite{romero15}, show a spread in mass ranging 
from $\sim$0.525\,M$_{\odot}$, at the top of the cooling sequence, to $\sim$0.875\,M$_{\odot}$, 
for NGC\,6397, and $\sim$0.520\,M$_{\odot}$ to $\sim$0.817\,M$_{\odot}$, for 47\,Tuc. The mass at the 
truncation for NGC\,6397 is significantly larger than the $\sim0.62\,M_{\odot}$ estimated by \cite{hansen07} 
for the same data. The models \cite{hansen07} used to determine the pre-white dwarf times attempted to take 
the effect of the metallicity into account. However, they considered as the white dwarf mass that at the first 
thermal pulse in the AGB. This approach, by not taking all the thermal pulses into account causes the major 
difference between the values of the final masses (Fig. \ref{masstwd}).

Another point that should be emphasised is that our models that best fit the data lead to masses at the 
top of the white dwarf cooling sequence of $\sim 0.525M_{\odot}$ for NGC\,6397 and $\sim 0.520M_{\odot}$ for 
47\,Tuc, that are consistent with the values found with spectroscopy of the brightest white dwarf stars in 
NGC\,6397, NGC\,6752 and M\,4 ($\overline{M}=0.53\pm0.03M_{\odot}$, \citealt{moehler04}; $\overline{M}=0.53\pm0.01M_{\odot}$, 
\citealt{kalirai09}).

\section{Conclusions}
\label{conclusions}

The first colour-magnitude diagrams of globular clusters down to the cutoff in the white dwarf cooling sequence exposed unprecedented features. One of these is the blue turn at magnitudes around M$_{\rm Bol}\sim14.7$ which was not well reproduced by single mass white dwarf models. 

The effect of different model assumptions on the mass spread of the white dwarf  stars and the importance of a consistent and detailed computation of their cooling  evolution becomes clear when we compare the ages of NGC\,6397 and 47\,Tuc determined by \cite{hansen13} (11.70\,Gyr and 9.90\,Gyr), with those obtained from our best-fit isochrone models (12.93\,Gyr and 10.95\,Gyr). Our absolute ages for both clusters are higher and, the age difference between NGC\,6397 and 47\,Tuc is 1.98$^{+0.44}_{-0.26}$\,Gyr. This difference is consistent with that determined by \cite{hansen13} of 2.0$\pm$0.5\,Gyr. However, if we wanted to consider the distance modulus and reddening of the main sequence, the absolute age of each cluster would change to 12.48\,Gyr (NGC\,6397) and 11.31\,Gyr (47\,Tuc), and the age deficit (1.17$_{-0.31}^{0.50}$\,Gyr) would be different than that by \citet{hansen13}.

Also, the absolute ages we determined are consistent with the formation epochs of metal-poor and metal-rich globular clusters determined by \citet{forbes15}. They inferred a mean formation epoch of globular clusters of 11.5$^{+0.6}_{-1.2}$\,Gyr for the metal-rich and, 12.2$^{+0.2}_{-0.3}$\,Gyr, for the metal-poor, if they all accreted from satellites, and  12.8$^{+0.2}_{-0.4}$\,Gyr,  if they are all formed within the main host galaxy. When they apply a similar method to the Milky Way they find 10.7\,Gyr for the metal-rich and 12.5-12.8\,Gyr for the metal-poor depending on whether they formed in accreted satellites or within the main host galaxy. Additionally, \citet{trenti15} derived an average of old globular clusters as 13.00$\pm$0.2\,Gyr through cosmological simulations, which, again, is consistent with the age we determine for NGC\,6397. 

Even though the red offset problem does not interfere in the fitting--because we keep the reddening as a fixed parameter--the models that best fit the data still face the red offset problem, i.e., change in the slope of white dwarf cooling sequence after M$_{\rm Bol}\sim14.6$, that shows a trend for the models to be bluer than the data, indicating that the construction of models of white dwarf stars still have significant aspects that must be addressed in the future. The red offset problem is possibly related to partial mixing of H and He in the atmosphere of white dwarf stars and/or the lack of a better physical description of the collision-induced absorption. By running some tests in our models, we notice that the effect of a very small mixing of hydrogen and helium in the atmosphere of the ultra-cool white dwarf stars correct the slope of the models. However this still is an ongoing investigation. Also, the red offset is in the same direction as that of mismatches observed when isochrone models are compared to the coolest main sequence stars, indicating that this might be a common problem for stars with very low temperatures, i. e., opacities for the coolest stars. Also, there are no models calculated with the newest collision-induced absorption opacities \citep{abel12}, so it is still and open question if they will help to solve the red offset problem.

We determined a very high value of $\alpha$ for 47\,Tuc, i.e., a bottom heavy present day local mass function in that portion of the cluster. \citet{heyl15} explained the lack of massive white dwarf stars near the centre of 47\,Tuc by the diffusion due to gravitational relaxation. The data we use in our analysis is very close to the centre of the cluster so, a bottom heavy present day local mass function is another indicator of the lack of massive white dwarf stars near the centre of 47\,Tuc. For NGC\,6397 we find an $\alpha$ consistent with a top-heavy type, in line with the analysis by \citet{richer08} and references therein. 

Another important feature highlighted in our analysis is the crystallisation peak, that is clearly visible in both, the data and the model, for NGC\,6397 at F606W$\sim$28, is not prominent for 47\,Tuc.  For this cluster, neither the models nor the data present a clear crystallisation peak.

\section*{Acknowledgements}
The authors would like to thank the anonymous referee for important suggestions that helped improve the manuscript. The authors would like to thank B.M.S. Hansen and H. B. Richer for their suggestions, for sending the photometric data and the artificial star tests of NGC\,6397. F.C. would like to thank J.S.Kalirai for sending the artificial star tests of 47\,Tuc. Partial financial support for this research comes from CNPq and PRONEX-FAPERGS/CNPq (Brazil). This work was supported in part by the NSERC Canada and by the Fund FRQ-NT (Qu\'ebec). M.H.M. and D.E.W. gratefully acknowledge the support of the NSF under grants AST-0909107 and AST-1312983. L.R.B. acknowledge PRIN-INAF 2012 funding under the project entitled: ``The M4 Core Project with Hubble Space Telescope''.

\bsp

\label{lastpage}

\end{document}